\documentclass[11pt]{article}
\usepackage{amssymb,amsmath}
\usepackage{paralist,cite}
\usepackage{graphicx}
\newtheorem{theorem}{\textbf{Theorem}}
\newtheorem{lemma}[theorem]{\textbf{Lemma}}

 \DeclareMathOperator{\wt}{wt}

\newcommand{\F}{\mathbf{F}}
\newcommand{\nix}[1]{}
\begin{document}
\title{Asymmetric Quantum Cyclic Codes}
\author{Salah A. Aly \\ Cairo University}
\date{}

\maketitle
%\begin{abstract}
\noindent {\bf  Abstract.}
It is recently conjectured    that phase-shift errors occur with high probability than qubit-flip errors, hence phase-shift errors are more
disturbing to quantum information than  qubit-flip
errors. This leads to construct  asymmetric quantum error-correcting codes (AQEC) to protect quantum information over asymmetric channels, $\Pr Z \geq \Pr X$. In this paper we present two generic
methods to derive asymmetric quantum cyclic codes  using the generator
polynomials and defining sets of classical cyclic codes. Consequently, the
methods allow us to construct several families of asymmetric quantum
BCH, RS, and RM codes. Finally, the methods are used to construct families of asymmetric subsystem codes.
%\end{abstract}

\bigskip

\noindent {\bf Construction of AQEC (Main Results).} The following theorem shows the CSS construction of asymmetric
quantum error control codes over $\F_q$.

%\medskip

\begin{theorem}[CSS AQEC]\label{lem:AQEC}
Let $C_1$ and $C_2$ be two classical codes with parameters
$[n,k_1,d_1]_q$ and $[n,k_2,d_2]_q$ respectively, and\\ $d_x=
\min\big\{\wt(C_{1} \backslash C_2^\perp), \wt(C_{2} \backslash C_{1
}^\perp)\big\}$, $d_z= \max\big\{\wt(C_{1} \backslash
C_2^\perp), \wt(C_{2} \backslash C_{1 }^\perp)\big\}$. If
  $C_2^\perp \subseteq C_1$, then
\begin{compactenum}[i)]
\item  there exists an AQEC with parameters $[[n,\dim C_1 -\dim
C_2^\perp,d_z/d_x]]_q$ that is $[[n,k_1+k_2-n,d_z/d_x]]_q$.
 Also, there
exists a QEC with parameters $[[n,k_1+k_2-n,d_x]]_q$.
\item there exists an asymmetric subsystem code with parameters $[[n,k_1+k_2-n-r,r,d_z/d_x]]_q$ for $0 \leq r \leq k_1+k_2-n$.
\end{compactenum} Furthermore, all constructed codes are pure to their minimum
distances.
\end{theorem}
%\begin{proof}
%\rnix{to be written}
%\end{proof}
Therefore, it is straightforward to derive asymmetric quantum
control codes from two classical codes as shown in
Theorem~\ref{lem:AQEC} as well as a subsystem code. Of course, one wishes to increase the values
of $d_z$ vers. $d_x$ for the same code length and dimension.
If the AQEC has minimum distances $d_z$ and $d_x$ with $d_z \geq d_x$, then the following Lemma holds.

%\medskip

\begin{lemma}
An $[[n,k,d_z/d_x]]_q$ asymmetric quantum code corrects all
qubit-flip errors up to $\lfloor (d_x-1)/2\rfloor$ and all
phase-shift errors up to $\lfloor (d_z-1)/2 \rfloor$.
\end{lemma}

\begin{theorem}
Let $C_1$ be a cyclic code with parameters $[n,k_1,d_1]_q$ and a
generator polynomial $g_1(x)$. Let $C_2^\perp$ be a cyclic code
defined by the polynomial $f(x)g_1(x)$ such that $b= \deg(f(x)) \geq
1$, then there exists AQEC with parameters
$[[n,2k_1-b-n,d_z/d_x]]_q$, s.t. $d_x=\min \{\wt(C_1 \backslash
C_2^\perp), \wt(C_2 \backslash C_1^\perp)\}$ and $d_z=\max \{\wt(C_1
\backslash C_2^\perp), \wt(C_2 \backslash C_1^\perp)\}$. Furthermore
the code can correct $\lfloor (d_x-1)/2 \rfloor$ qubit-flip errors
and  $\lfloor (d_z-1)/2 \rfloor$ phase-shift errors.
\end{theorem}

\begin{theorem}\label{lem:cyclic-subsysI}
Let $C_1$ be a $k$-dimensional cyclic code of length $n$ over
$\F_q$. Let $T_{C_1}$ and $T_{C_1^\perp}$ respectively denote the
defining sets of $C_1$ and $C_1^\perp$. If $T$ is a subset of
$T_{C_1^\perp} \setminus T_{C_1}$ that is the union of cyclotomic
cosets, then one can define a cyclic code $C_2$ of length $n$ over
$\F_q$ by the defining set $T_{C_2}= T_{C_1^\perp} \setminus (T \cup
T^{-1})$. If $b=|T\cup T^{-1}|$ is in the range $0\le b< 2k-n$ then
there exists asymmetric quantum code with parameters
$[[n,2k-b-n,d_z/d_x]]_q,$ where $d_x= \min \{ \wt(C_2 \setminus C_1^\perp),\wt(C_1 \setminus C_2^\perp) \}$
and $d_z= \max \{ \wt(C_2 \setminus C_1^\perp),\wt(C_1 \setminus
C_2^\perp) \}$.
\end{theorem}

The usefulness of the previous theorem is that one can directly
derive asymmetric quantum codes from the set of roots (defining set)
of a classical cyclic code. We also notice that the integer $b$ represents a
size of a cyclotomic coset (set of roots), in other words, it does
not represent one root in $T_{C_1^\perp}$.

\medskip

\noindent
{\bf AQEC and Connection with Subsystem
Codes.}\label{sec:AQEC-subsystem}
We establish the connection between AQEC and
subsystem codes. Furthermore we derive a larger class of quantum
codes called asymmetric subsystem codes (ASSC).

\begin{theorem}[ASSC Euclidean Construction]\label{lem:css-Euclidean-subsys}
If $C_1$ is a $k_1$-dimensional $\F_q$-linear code of length $n$ that
has a $k_2$-dimensional subcode $C_2=C_1\cap C_1^\perp$ and
$k_1+k_2<n$, then there exist
\begin{eqnarray}
[[n,n-(k_1+k_2),k_1-k_2,d_z/d_x]]_q,\nonumber \\  \nonumber  [[n,k_1-k_2,n-(k_1+k_2),d_z/d_x]]_q  \end{eqnarray}
subsystem codes, where $d_z=\max\{\wt(C_2^\perp\setminus C_1),\wt(C_1^\perp\setminus C_2)\}$ and $d_x=\min\{\wt(C_2^\perp\setminus C_1),\wt(C_1^\perp\setminus C_2)\}$.
\end{theorem}

 From this
result, we can see that any two classical codes $C_1$ and $C_2$ such
that $C_2=C_1 \cap C_1^\perp \subseteq C_2^\perp$, in which they can
be used to construct a subsystem code (SSC), can also be used to
construct asymmetric quantum code (AQEC). Asymmetric subsystem codes
(ASSC) are a much larger class than  the class of symmetric subsystem
codes, in which the quantum errors occur with different
probabilities in the former one and have equal probabilities in the
later one.

The interested readers may look at the previous work in constructions of asymmetric quantum error control codes (AQEC)~\cite{evans07,ioffe07,stephens07,steane97}. The CSS construction of QEC is well explained in~\cite{calderbank98,gottesman97}, and in the list of references presented in~\cite{aly08phd}.

\end{document}